\documentclass[sigconf]{acmart}
\usepackage{algorithm}
\usepackage{algpseudocode}
\usepackage{tabu}
\usepackage{array}
\usepackage{tikz}
\usetikzlibrary{matrix, positioning}
\AtBeginDocument{%
  }

\setcopyright{acmlicensed}
\copyrightyear{2018}
\acmYear{2018}
\acmDOI{XXXXXXX.XXXXXXX}
\acmConference[Conference acronym 'XX]{Companion Proceedings of the ACM Web Conference 2026}{June 03--05,
  2018}{Woodstock, NY}
\acmISBN{978-1-4503-XXXX-X/2018/06}

\begin{document}

\title{RIA: A Ranking-Infused Approach for Optimized listwise CTR Prediction}

\author{Guoxiao Zhang}
\authornote{Both authors contributed equally to this research.}
\email{zhangguoxiao@meituan.com}
\orcid{0009-0004-2628-7094}
\author{Tan Qu}
\authornotemark[1]
\email{qutan@meituan.com}
\affiliation{%
  \institution{Meituan}
  \city{Beijing}
  \country{China}
}
\author{Ao Li}
\authornotemark[1]
\affiliation{%
  \institution{Meituan}
  \city{Beijing}
  \country{China}}
\email{liao27@meituan.com}
\author{Donglin Ni}
\affiliation{%
  \institution{Beijing University of Posts and Telecommunications}
  \city{Beijing}
  \country{China}}
\email{nidl@bupt.edu.cn}

\author{Qianlong Xie}
\affiliation{%
  \institution{Meituan}
  \city{Beijing}
  \country{China}}
  \email{xieqianlong@meituan.com}

\author{Xingxing Wang}
\affiliation{%
  \institution{Meituan}
  \city{Beijing}
  \country{China}}
  \email{wangxingxing04@meituan.com}

\renewcommand{\shortauthors}{Guoxiao Zhang, Tan Qu, Ao Li, Qiang Liu, and Xingxing Wang}

\begin{abstract}
Reranking improves recommendation quality by modeling item interactions. However, existing methods often decouple ranking and reranking, leading to weak listwise evaluation models that suffer from combinatorial sparsity and limited representational power under strict latency constraints. In this paper, we propose \textbf{RIA} (Ranking-Infused Architecture), a unified, end-to-end framework that seamlessly integrates pointwise and listwise evaluation. \textbf{RIA} introduces four key components: (1) \textbf{the User and Candidate Dual-Transformer (UCDT)} for fine-grained user-item-context modeling; (2) \textbf{the Context-aware User History and Target (CUHT)} module for position-sensitive preference learning; (3) \textbf{the Listwise Multi-HSTU (LMH)} module to capture hierarchical item dependencies; and (4) \textbf{the Embedding Cache (EC)} module to bridge efficiency and effectiveness during inference. By sharing representations across ranking and reranking, \textbf{RIA} enables rich contextual knowledge transfer while maintaining low latency. Extensive experiments show that \textbf{RIA} outperforms state-of-the-art models on both public and industrial datasets, achieving significant gains in AUC and LogLoss. Deployed in Meituan's advertising system, \textbf{RIA} yields a +1.69\% improvement in Click-Through Rate (CTR) and a +4.54\% increase in Cost Per Mille (CPM) in online A/B tests.
\end{abstract}



\maketitle

\section{Introduction}
\begin{figure}
    \centering
    \includegraphics[width=\linewidth]{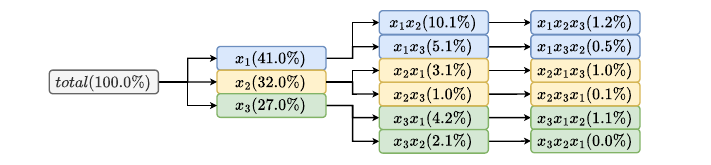}
    \caption{
    An illustration of \emph{combinatorial sparsity}.
  }
  \label{fig:combinatorial-sparsity}
\end{figure}
Reranking aims to optimize the final item list by modeling interactions among candidates, enabling listwise utility maximization. While early generator-only methods~\cite{ai2018learning,pei2019personalized,xi2022contextawarererankingutilitymaximization} suffer from the \emph{evaluation-before-reranking} dilemma~\cite{xi2022contextawarererankingutilitymaximization}, recent evaluation-based approaches~\cite{feng2021revisit, shi2023pier, yang2025comprehensive} adopt a two-stage paradigm: generating candidate lists followed by listwise scoring. However, these methods primarily focus on list generation, leaving the listwise evaluation model under-explored. More recently, the evaluation-only method YOLOR~\cite{wang2025you} bypasses list generation but suffers from two limitations: it implicitly assumes conditional independence between position and context, and its computational cost limits scalability to millions or more candidate lists.

A key challenge for listwise evaluation is \emph{combinatorial sparsity}, the exponential decay in co-exposure frequency as list size increases. As illustrated in Figure~\ref{fig:combinatorial-sparsity}, while individual items may be frequently exposed, their joint occurrences (e.g., triplets) are extremely rare, making it difficult to learn robust contextual interactions from data.
Compounding this issue, those listwise evaluation model are architecturally decoupled from the pointwise evaluation models during the initial ranking stage. Designed under strict latency constraints\footnote{The rerank stage processes fewer candidates than the rank stage, so the recommendation pipeline allocates less time consumption to it.}, they tend to be simpler and less expressive, creating a representational gap that hinders knowledge transfer and limits modeling capacity.

To address these limitations, we propose \textbf{RIA} (Ranking-Infused Architecture), a unified evaluation-based framework that seamlessly integrates pointwise and listwise evaluation models into a single end-to-end pipeline. \textbf{RIA} consists of four components: (1) the \textbf{User and Candidate Dual-Transformer (UCDT)} for fine-grained candidate-context interaction modeling; (2) the \textbf{Context-aware User History and Target (CUHT)} module for position-aware preference learning; (3) the \textbf{Listwise Multi-HSTU (LMH)} module for hierarchical cross-item dependency modeling; (4) the \textbf{Embedding Cache (EC)} module to enhance inference efficiency.

Our contributions are threefold: (i) We present \textbf{RIA}, the first framework to unify pointwise and listwise evaluation under a single end-to-end paradigm; (ii) We design a powerful yet efficient listwise evaluation model with novel modules, which outperforms state-of-the-art models on both public and industrial datasets; (iii) We deploy \textbf{RIA} on Meituan's advertising system, where online A/B tests show a +1.69\% gain in CTR and +4.54\% in CPM, validating its real-world effectiveness.

\begin{figure*}[t]
    \centering
    \scalebox{1.0}{
    \includegraphics[width=\linewidth]{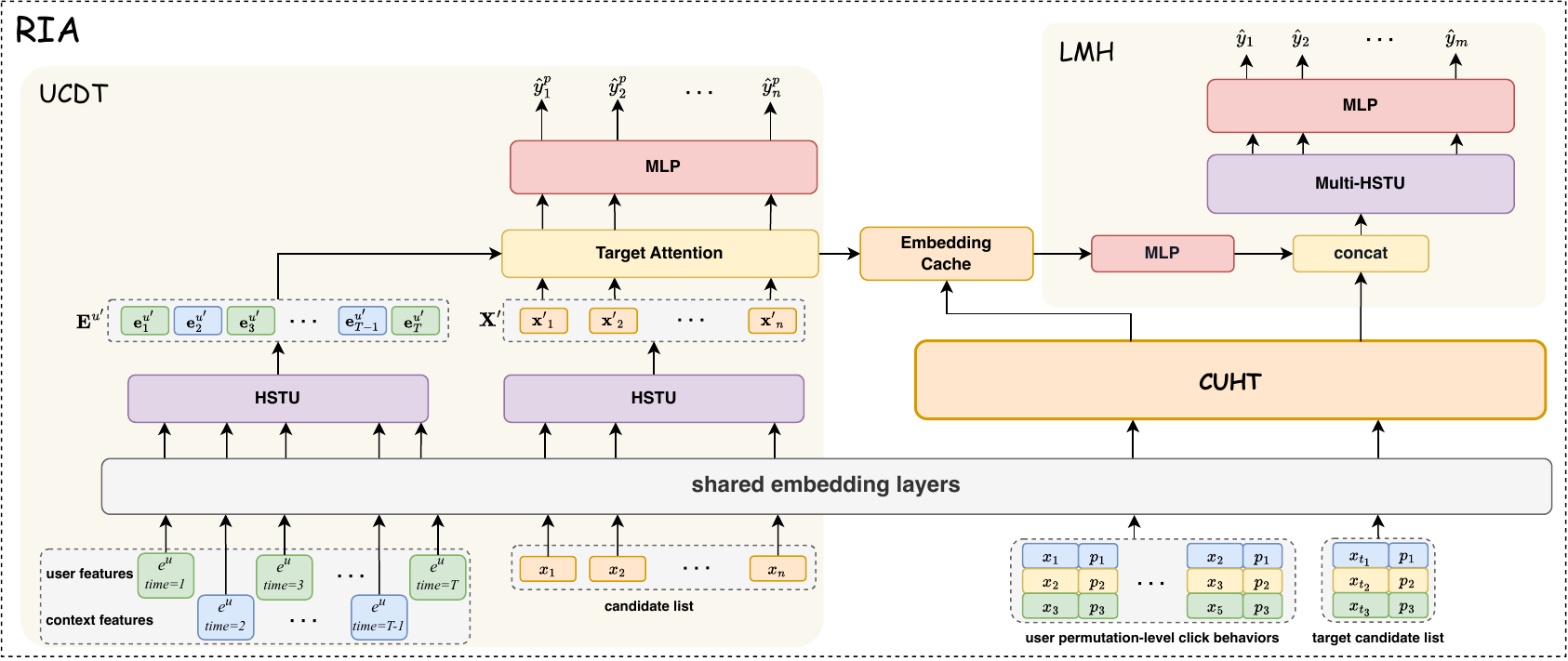}
    }
    \caption{{Overview of our proposed method ($m=3$).}}
    \label{fig:pipeline}
    \vspace{0mm}
    \Description{Overview diagram of the proposed method.}
\end{figure*}
\section{Methodology}


Our framework, illustrated in Figure~\ref{fig:pipeline}, seamlessly integrates pointwise and listwise evaluation through four synergistic modules: \textbf{UCDT}, \textbf{CUHT}, \textbf{LMH}, and \textbf{EC}. For each user $ u $, given a candidate list $ X = \{x_1, \dots, x_n\} $ from the ranking stage and a set of generated target candidate lists $ \mathcal{P} $ by generative models, where each target candidate list $ \mathbf{P} \in \mathcal{P} $ contains $ m $ items ($ m \leq n $), the goal is to select the optimal list $ \mathbf{P}^* $ that maximizes the expected listwise reward:
\begin{equation}
    \mathbf{P}^*  = \arg\max_{\mathbf{P} \in \mathcal{P}} R(u, \mathbf{P}), \quad \text{where} \quad R(u, \mathbf{P}) = \sum_{i=1}^{m} r(u, x_i).
\end{equation}
Here, $ r(u, x_i) $ denotes the reward for item $ x_i $, and $ R(u, \mathbf{P}) $ represents the cumulative reward of the entire list $ \mathbf{P} $.

\begin{figure}[ht]
    \centering
    \includegraphics[width=0.95\linewidth]{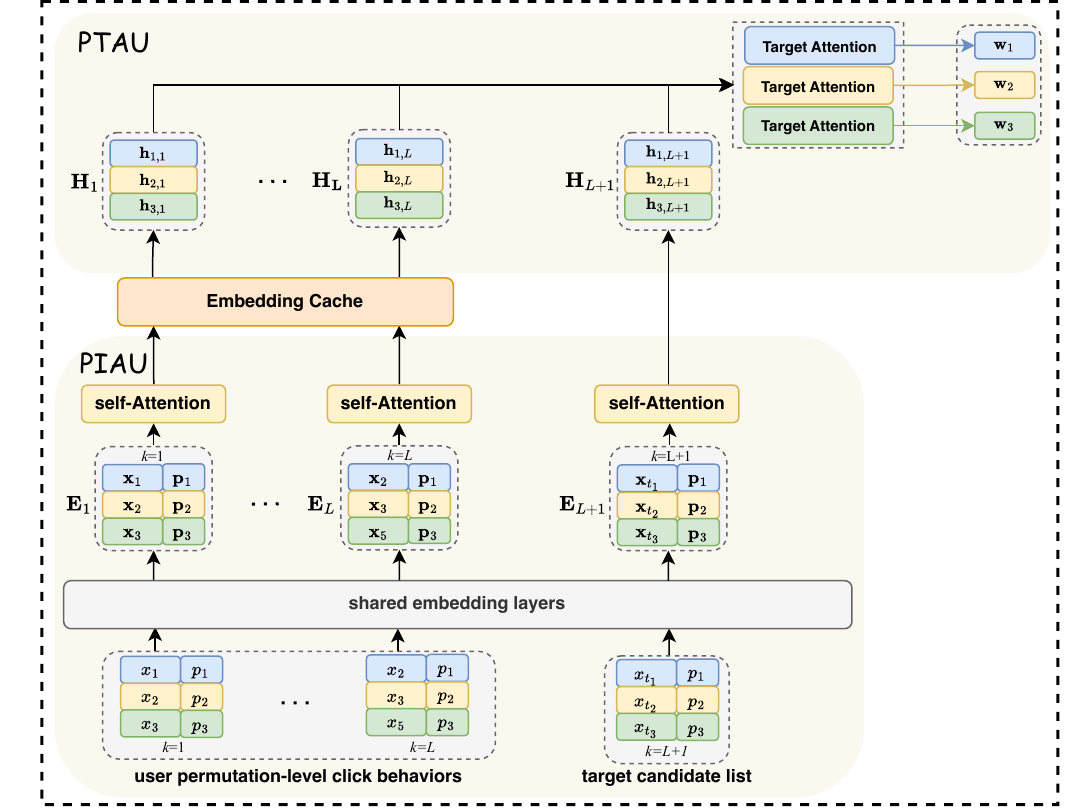}
    \caption{Architecture of CUHT ($m=3$).}
    \label{fig:UHTC}
\end{figure}
\subsection{User and Candidate Dual-Transformer (UCDT)}
The \textbf{UCDT} module captures fine-grained interactions between candidates and user context. Following the approach of ~\cite{zhai2024actions}, we merge user profile features, context features into the time series composed with user pointwise behaviors\footnote{which ignore the contextual items within each session.}, which is called user-context features (defined as $\mathbf{e}^{u}$). Let $\mathbf{X} \in \mathbb{R}^{n \times D}$ and $\mathbf{E}^{u} \in \mathbb{R}^{T \times D}$ denote embeddings of candidate list and user-context features, where $n$ is the number of candidate items, $D$ is the dimension of the embedding layer and ${T}$ is the length of the time series. 

Both $\mathbf{E}^u$ and $\mathbf{X}$ are encoded using Hierarchical Sequential Transduction Units (HSTU)~\cite{zhai2024actions}:
\begin{align}
    \mathbf{X}' &= \text{HSTU}(\mathbf{X}), \\
    \mathbf{E}^{u'} &= \text{HSTU}(\mathbf{E}^u).
\end{align}

We then apply a target attention mechanism~\cite{zhou2018deep} to model the interaction between each candidate item and the user-context features:
\begin{equation}
    \mathbf{x}_i'' = \text{Attention}(\mathbf{x}_i', \{\mathbf{e}_j^{u'}\}_{j=1}^{T}), \quad i =1,\dots,n.
    \label{eq:4}
\end{equation}
where $\mathbf{x}_i' \in \mathbb{R}^D$ is the $i$-th candidate item in $\mathbf{X'}$, $\mathbf{e}_j^{u'} \in \mathbb{R}^D$ is the $j$-th user-context feature in $\mathbf{E^{u'}}$.

The pointwise CTR prediction ${\hat{y}^p_i}$ is:

\begin{equation}
    \hat{y}^p_i = \sigma\left( \mathrm{MLP}(\mathbf{x}_i'') \right),\quad i = 1,\dots,n.
\end{equation}
where $\sigma$ is the sigmoid function. 
The corresponding loss is calculated as binary cross-entropy loss defined as 
$\mathcal{L}_1$.

\subsection{Context-aware User History and Target (CUHT)}
In user behavior modeling, \textbf{UCDT} treats each action in isolation and ignores the contextual items within the same session.  
To address this limitation, we propose \textbf{CUHT}, as illustrated in Figure~\ref{fig:UHTC}, which consists of two core components: (1)  \textit{Page-level Inner Attention Unit} (PIAU) that models intra-session context via self-attention over user permutation-level historical behaviors\footnote{which incorporate session-level contextual information from past user interactions.} and the target candidate list; and (2) \textit{Position-aware Target Attention Unit} (PTAU) that explicitly models the interaction between the target candidate list and the permutation-level historical behavior at each session position.  
 
\subsubsection{PIAU}
Using shared embedding layers from \textbf{UCDT}, we denote $\mathbf{E}_k \in \mathbb{R}^{m \times D'}$ and $\mathbf{E}{_{L+1}} \in \mathbb{R}^{m \times D'}$  as the $k$-th permutation embeddings of user permutation-level historical behaviors with length $L$ and the embeddings of the target candidate list\footnote{The position embeddings are included.}, where $D'$ is the sum of $D$  and the position embedding size. PIAU applies a parameter-share self-attention layer to calculate the mutual influence of different items and output corresponding matrix $\mathbf{H}_k$, as follows:
\begin{equation}
    \mathbf{H}_k = \mathrm{selfAttention}\left(\mathbf{E}{_k}\right), \quad  k=1,\dots,L+1.
\end{equation}

\subsubsection{PTAU}
Specifically, for a given position $o$ in the session, PTAU computes target attention between the $o$-th item in the target candidate list and all items in the permutation-level historical behaviors at position $o$:
\begin{equation}
    \mathbf{w}_o = \text{Attention}(\mathbf{h}_{o,L+1}, \{\mathbf{h}_{o,k}\}_{k=1}^{L}), \quad o =1,\dots,m.
\end{equation}
where $\mathbf{h}_{o,k}$ represents the final representation at the $o$-th position in $\mathbf{H}_k$, and $\mathbf{h}_{o,L+1}$ represents the final representation in $\mathbf{H}_{L+1}$.

\subsection{Listwise Multi-HSTU (LMH)}
\textbf{LMH} models hierarchical list-level dependencies. It first transforms middle representations  $\mathbf{x}_o''$ from Equation~\eqref{eq:4} via a MLP adaptor:
\begin{equation}
    \mathbf{t}_o = \mathrm{MLP}(\mathbf{x}_o''), \quad o =1,\dots,m.
\end{equation}

The target list representation is built by stacking HSTU layers( with layer num as $I$) over concatenated $\mathbf{t}_o$ and $\mathbf{w}_o$:

\begin{equation}
\mathbf{M}_i = [\mathbf{m}_{i,1},\dots,\mathbf{m}_{i,m}] =
\begin{cases}
\mathrm{HSTU}\left( [\mathbf{t}_1 \parallel \mathbf{w}_1, \dots, \mathbf{t}_m \parallel \mathbf{w}_m] \right), & i = 1, \\[0.8em]
\mathrm{HSTU}(\mathbf{M}_{i-1}), \quad i = 2,\dots,I.
\end{cases}
\end{equation}


where $||$ denotes concatenation and $\mathbf{m}_{i,o}$ is the features of the $o$-th position in $\mathbf{M}_{i}$.
Then, the listwise pCTR of the $o$-th item in the target list is predicted as follows:
\begin{equation}
    \hat{y}_o = \sigma\left( \mathrm{MLP}\left( \mathbf{m}_{I,o} \right) \right),\quad o = 1,\dots,m.
    \label{eq:prediction}
\end{equation}
Subsequently, the listwise loss is binary cross-entropy loss defined as $\mathcal{L}_2$. Finally, the total loss is:
\begin{equation}
    \mathcal{L} = \mathcal{L}_1 + \mathcal{L}_2.
\end{equation}
\subsection{Embedding Cache (EC)}
Due to strict time constraints, such a complex method cannot be applied directly to the reranking module. Therefore, we further design an EC module to enable efficient reuse. During ranking, We pre-compute the item-level representation $\mathbf{x}_i''$ and the user permutation-level click behaviors representation $\mathbf{H}_k$ through the aforementioned \textbf{UCDT} and PIAU in \textbf{CUHT} modules, respectively, as shown in Figure~\ref{fig:pipeline} and Figure~\ref{fig:UHTC}. 

\section{EXPERIMENTS}

\subsection{Experimental Setup}

\subsubsection{Datasets.}
We evaluate our framework on both a public benchmark and a large-scale industrial dataset. The public \textbf{Avito} dataset contains user search and ad interaction logs. The industrial \textbf{Meituan} dataset is collected from Meituan's advertising system, reflecting real-world user behavior in a large-scale local services ecosystem. Key statistics are summarized in Table~\ref{tab:dataset_statistics}.

\begin{table}[h]
\centering
\caption{Dataset statistics.}
\label{tab:dataset_statistics}
\begin{tabular}{lccc}
\toprule
\textbf{Dataset} & \textbf{\#Requests} & \textbf{\#Users} & \textbf{\#Items} \\
\midrule
Avito  & 53,562,269 & 1,324,103 & 23,562,269 \\
Meituan & 88,279,996 & 24,074,754 & 9,190,395 \\
\bottomrule
\end{tabular}
\end{table}

\subsubsection{Baselines.}
We compare our model against representative pointwise and listwise CTR models:
\begin{itemize}
    \item \textbf{PRM}~\cite{pei2019personalized}: Uses self-attention to model item dependencies in reranking.
    \item \textbf{OCPM}~\cite{shi2023pier}:  Listwise model with omnidirectional attention and context-aware prediction.
    \item \textbf{YOLOR}~\cite{wang2025you}: State-of-the-art listwise model with multi-scaling context information.
    \item \textbf{RIA\_small}: Our model with 1 HSTU layer in the LMH module.
    \item \textbf{RIA\_big}: Our model with 8 HSTU layers.
\end{itemize}

\subsubsection{Evaluation Metrics.}
We report \textbf{AUC} and \textbf{LogLoss} for offline evaluation, following standard practice in CTR prediction.

\subsection{Overall Performance}
Table~\ref{tab:performance-comparison} presents the offline performance comparison. Our proposed \textbf{RIA} consistently outperforms all baselines on both datasets.

On the Avito dataset, \textbf{RIA\_small} improves AUC by +0.40\% over YOLOR, while \textbf{RIA\_big} achieves a +0.85\% gain. On the larger Meituan dataset, the improvements of AUC are even more pronounced: +0.31\% (\textbf{RIA}\_small) and +0.96\% (\textbf{RIA}\_big). These results validate the effectiveness of our unified architecture in capturing both fine-grained interactions and listwise dependencies.

\begin{table}[ht]
    \centering
    \caption{Offline performance comparison.}
    \label{tab:performance-comparison}
    \begin{tabular}{lccccc}
        \toprule
        \textbf{Model} & \multicolumn{2}{c}{\textbf{Avito}} & \multicolumn{2}{c}{\textbf{Meituan}} \\
        \cmidrule(lr){2-3} \cmidrule(lr){4-5}
        & \textbf{AUC} & \textbf{LogLoss} & \textbf{AUC} & \textbf{LogLoss} \\
        \midrule
        PRM    & 0.7131 & 0.0481 & 0.6541 & 0.1614 \\
        OCPM   & 0.7320 & 0.0471 & 0.6624 & 0.1596 \\
        YOLOR   & 0.7340 & 0.0470 & 0.6634 & 0.1595 \\
        \midrule
        \textbf{RIA\_small} & \textbf{0.7380} & \textbf{0.0468} & \textbf{0.6665} & \textbf{0.1592} \\
        \textbf{RIA\_big}   & \textbf{0.7425} & \textbf{0.0456} & \textbf{0.6730} & \textbf{0.1483} \\
        \bottomrule
    \end{tabular}
\end{table}

\subsection{Scaling Behavior of the LMH Module}
\begin{figure}[ht]
    \centering
    \includegraphics[width=0.8\linewidth]{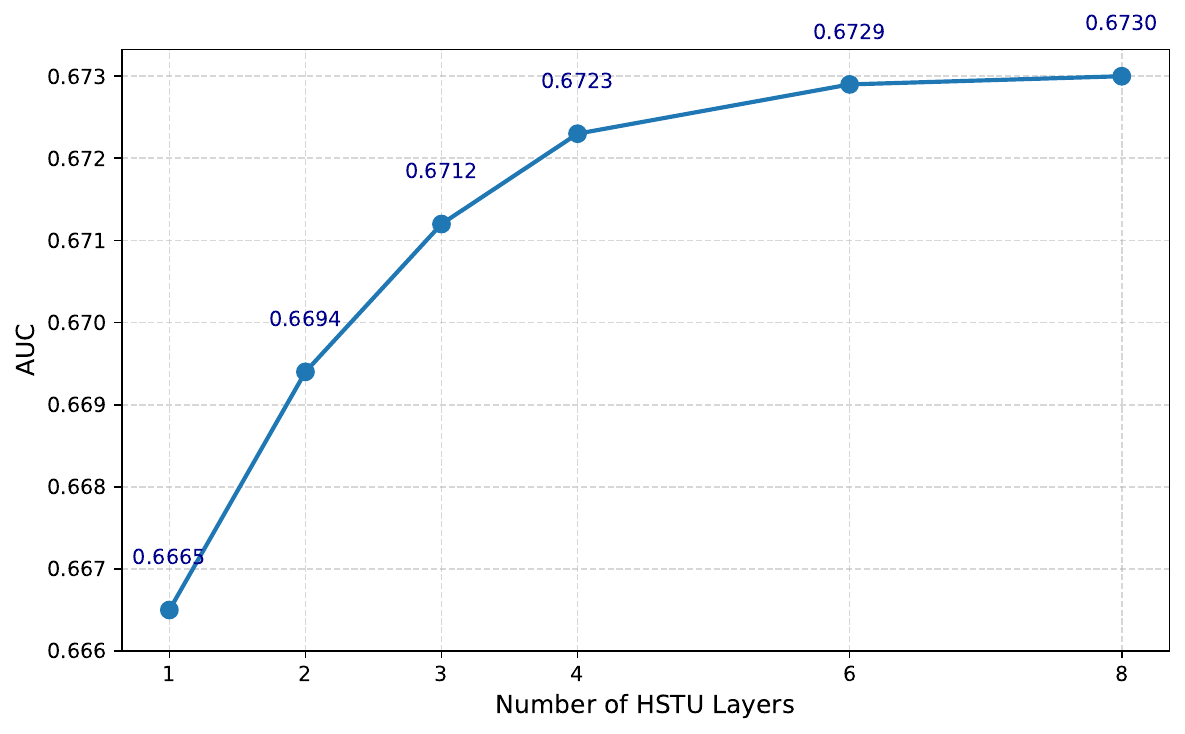}
    \caption{AUC vs. Number of HSTU layers.}
    \label{fig:auc_vs_layers}
\end{figure}

We analyze the impact of architectural depth in the \textbf{LMH} module by varying the number of HSTU layers on Meituan dataset. As shown in Figure~\ref{fig:auc_vs_layers}, AUC increases monotonically with depth on the Meituan dataset, rising from 0.6665 (1 layer) to 0.6730 (8 layers). This consistent improvement suggests a \emph{scaling law} in listwise modeling.

\subsection{Online Deployment and A/B Test Results}
To enable real-time inference, we deploy a hybrid optimization strategy combining pre-computation and hierarchical caching, inspired by MSD~\cite{zhang2025balancing}. Specifically, the \textbf{UCDT} module and PIAU component are computed in parallel with the ranking pipeline. Embeddings for top-$n$ candidates and recent user behavior sequences are pre-computed and cached in Redis, significantly reducing online latency.

We conducted an A/B test on Meituan's advertising system from September 6 to 14, 2025. As shown in Table~\ref{tab:ab_test_results}, \textbf{RIA\_small} achieves a +1.69\% increase in CTR and +4.54\% in CPM over the baseline\footnote{Since YOLOR is not suited to our scenario, we adopt OCPM as the baseline.} while keeping latency nearly unchanged (26.1 ms to 28.2 ms). \textbf{RIA\_big} achieves even greater performance gains (+2.11\% CTR, +5.83\% CPM). These results demonstrate the practical effectiveness and business value of our framework.

\begin{table}[ht]
    \centering
    \caption{Online A/B testing results.}
    \label{tab:ab_test_results}
    \begin{tabular}{lccc}
        \toprule
        \textbf{Method} & \textbf{CTR Gain} & \textbf{CPM Gain} & \textbf{Latency} \\
        \midrule
        Baseline         & - & - & 26.1 ms \\
        \textbf{RIA\_small} & \textbf{+1.69\%} & \textbf{+4.54\%} & 28.2 ms \\
        \textbf{RIA\_big}   & \textbf{+2.11\%} & \textbf{+5.83\%} & 36.7 ms \\
        \bottomrule
    \end{tabular}
\end{table}

\section{Conclusion}
In this paper, we propose \textbf{RIA}, a unified framework for end-to-end listwise CTR prediction that seamlessly integrates ranking and reranking. Extensive experiments show that \textbf{RIA} outperforms state-of-the-art baselines, achieving significant AUC gains on both public and industrial datasets. Our work bridges the gap between ranking and reranking through a single, trainable architecture, offering both performance gains and deployment efficiency. We hope this unified paradigm inspires future research on holistic, context-aware ranking systems in large-scale recommendation scenarios.

\bibliographystyle{ACM-Reference-Format}
\balance
\bibliography{rank_listwise}
\end{document}